\documentclass{iopart}

\usepackage{amssymb}
\usepackage{graphicx}

\newcommand{\half}{\textstyle\frac{1}{2}}

\begin{document}

\title[Fractal dimension of transport coefficients\ldots]{Fractal
dimension of transport coefficients\\ in a deterministic dynamical system}

\author{Zbigniew Koza}

\address{
             Institute of Theoretical Physics, University of
             Wroc{\l}aw, pl.\ Maxa Borna 9, 50204 Wroc{\l}aw,
             Poland
           }
\ead{zkoza@ift.uni.wroc.pl}

\date{\today}

\begin{abstract}
In many low-dimensional dynamical systems transport coefficients are very
irregular, perhaps even fractal functions of control parameters. To analyse
this phenomenon we study a dynamical system defined by a piece-wise linear map
and investigate the dependence of transport coefficients on the slope of the
map. We present analytical arguments, supported by numerical calculations,
showing that both the Minkowski-Bouligand and Hausdorff fractal dimension of the graphs of these
functions is 1 with a logarithmic correction, and find that the exponent
$\gamma$ controlling this correction is bounded from above by 1 or 2,
depending on some detailed properties of the system. Using numerical
techniques we show local self-similarity of the graphs. The local
self-similarity scaling transformations turn out to depend (irregularly) on
the values of the system control parameters.
\end{abstract}

\submitto{\JPA}


\section{Introduction}
\label{Introduction}

Various transport coefficients associated with a particle moving in
low-dimensional periodic array of scatterers are often very irregular
functions of control parameters. For example, the diffusion coefficient of a
particle moving in the flower-shaped billiard (a 2D Lorentz gas with
periodically distributed scatterers of six-petal flower shape) was found to be
a very irregular function of the petal curvature \cite{Har02}, and the
conductivity of the classical Lorentz gas (with disc scatterers) varies
rapidly with the applied field \cite{Lloyd95}.  The reaction rate and the
diffusion coefficient in the reaction-diffusion multibaker model also turned
out to be irregular, fractal-like functions of the map's shift
\cite{GaspardKlages98}. Similar behaviour  was found for the diffusion
coefficient of a particle thrown onto a periodically corrugated floor and
subject to various types of external force, including Hamiltonian
\cite{Harayama01} and dissipative \cite{Matyas03,KlagesHab} systems. In
chaotic systems with two degrees of freedom the modes governing the relaxation
to the thermodynamic equilibrium form a fractal structure with a nontrivial
fractal dimension which can be related to the Lapunov exponent and the
diffusion coefficient of the system \cite{Gilbert01,Gaspard01,Claus02}.
Another example is provided by a deterministic dynamical system with dynamics
defined by iteration of a one-dimensional map: the mean velocity and the drift
coefficient in this system were shown to be irregular, nowhere differentiable
functions of the system control parameters
\cite{G-K,KlagesKlauss03,Klages95,KlagesPRE99,KlagesPHD}.

It is natural to ask what the origin of these ``irregularities'' is, how to
describe them quantitatively, and to what extent they are universal. Answering
all these questions is far from trivial. One of the reasons why irregular
behaviour of transport coefficients is so difficult to investigate is that
until very recently our knowledge about this phenomenon was based mainly on
numerical simulations. Although several different techniques of calculating
these quantities were developed, e.g.\ the transition matrix technique
combined with the escape-rate formalism
\cite{KlagesHab,KlagesPRE99,KlagesPHD,DorfmanBook,Gaspard90,Vollmer03,Gaspard92},
the Green-Kubo formula \cite{KlagesHab,KlagesPHD,DorfmanBook}, or the
periodic-orbit formalism \cite{DorfmanBook,CvitanovicWWW}, they all lead to
complicated and time-consuming numerical calculations of limited accuracy.
Moreover, quite often these techniques are applicable only for some special
values of the control parameters. What is worse, as a model becomes more realistic,
the results are getting less accurate.
 For this
reason the most precise data have been obtained for very simple,
one-dimensional ``toy models'', while the question whether a more realistic 2D
Lorentz gas (with scatterers of disc shape) exhibits any irregularities in the
transport coefficient as functions of the disc radius is still a subject of
controversy \cite{KlagesHab,KlagesDellago}.

This situation improved considerably when Groeneveld and Klages \cite{G-K} gave
exact formulae for the transport coefficients in a simple one-dimensional
deterministic dynamical system, introduced by Grossmann and Fujisaka
\cite{Grossman82}, and defined by a piece-wise linear map with two control
parameters: the slope and the bias. Their solution has two important features.
First, it is complete, i.e.\ applicable for any (physically meaningful) values
of the control parameters. Second, it reduces the problem of determining the
transport coefficients to finding the sum of a quickly converging series. For
the first time we thus have a model exhibiting irregular behaviour of the
transport coefficients, for which a solution permitting an efficient
numerical implementation is available.

Klages and Klau\ss\ \cite{KlagesKlauss03} have recently used this exact
solution to examine in detail irregular dependency of the drift velocity and
diffusion coefficient on the control parameters of the system. Their aim was
to verify an earlier hypothesis \cite{Klages95,KlagesPRE99} that the graphs of
these quantities as functions of the map slope are so irregular that actually
they form fractals. Using two numerical techniques: the box counting and the
autocorrelation function methods, they found that the local fractal dimensions
of these graphs are well-defined on small finite subintervals,
but highly irregular functions of the
control parameters. In other words they found that these graphs cannot  be
described with a single fractal dimension, but rather by a set of quickly
varying local fractal dimensions. Taking this into account they put forward a
hypothesis that the local fractal dimensions of these graphs as functions of
the slope are fractal themselves.

However, our recent numerical calculations \cite{Koza04} suggest that the
Minkowski-Bouligand fractal dimension $\Delta$ of the graph of the diffusion
coefficient as a function of the slope is equal 1, and that the convergence to
this limit is slowed down by a logarithmic correction. Thanks to this
logarithmic term the curve is a fractal. We also proposed a conjecture that the
exponent controlling this correction depends on the slope of the map and
equals either 1 or 2, depending on existence and detailed properties of a
Markov partition. That $\Delta =1$ seems to be at odds with the above-cited
results of \cite{KlagesKlauss03}, because the Minkowski-Bouligand dimension,
if exists, is equivalent to the box-counting dimension \cite{Tricot}. However,
one should note that while in our studies we focused on the point-wise
dimension calculated only for those system control parameters that correspond
to finite Markov partitions, Klages and Klau\ss\ computed the box-counting
dimension on intervals of finite length, and these two values need not be the
same. The aim of our present study is to support our earlier findings
analytically. We also present new numerical results that confirm these results.

The structure of the paper is as follows. Section \ref{sec:Formalism}
introduces the mathematical formalism. This includes the definition of the
model, a brief description of the Minkowski-Bouligand dimension, the
``oscillation'' method of calculating it for continuous curves, and precise
formulation of our hypothesis about the logarithmic corrections. Our main
results are contained in \sref{sec:case.b=0}, which presents a detailed
calculation of the fractal dimension of the graph of the diffusion coefficient
as a function of the slope for the case of zero bias. It also includes the
derivation of the upper bound for the exponent controlling the logarithmic
corrections. Section \ref{sec:b.neq.0} contains a brief discussion of how our
results can be generalized for the case of arbitrary bias and for other
transport coefficients. Section \ref{sec:Numerical} contains numerical
results. These include analysis of the factors determining the value of the
logarithmic correction as well as a study of self-similarity of the graphs.
Finally, \sref{sec:Conclusions} is devoted to discussion of results and
conclusions.


\section{Basic formalism}
\label{sec:Formalism}

\subsection{The model}

\label{sub:model}

Following Groeneveld and Klages \cite{G-K} we investigate a dynamical system
with the equation of motion defined by a one-dimensional map $M_{a,b}\colon
\mathbf{R} \to \mathbf{R}$ parameterized by some real numbers $a$ and $b$
\begin{equation}
  x_{n+1} = M_{a,b}(x_n)
\end{equation}
where $x_n$ represents position of a particle and $n$ is a discrete-time
variable.
The map is a piece-wise linear function given by
\begin{equation}
  \label{eq:defM.1}
    M_{a,b}(x) = \left\{\begin{array}{l@{\quad:\quad}l} ax+b & x\in [-\half,\half)\\
                                          M_{a,b}(x + n) - n &  x\in\mathbf{R}
                                          \end{array} \right.
\end{equation}
where $n$ is an arbitrary integer, and the slope $a>1$ and the bias $b \in
\mathbf{R}$ are the control parameters of the system (detailed analysis  of the
system symmetries \cite{G-K} leads to the conclusion that the values of $b$
can be restricted to  $-\half \le b < \half$). The interval $[-\half,\half)$
will be called ``the fundamental interval'' and denoted by $I_0^-$. However,
instead of $I_0^-$ one can choose $I_0^+ = (-\half,\half]$ as the fundamental
interval in (\ref{eq:defM.1}). It turns out that the transport coefficients
(defined below) are the same
 whether we choose $I_0^-$ or $I_0^+$ as the fundamental interval.  This
equivalence has been employed in the derivation of the explicit forms of the
transport coefficients \cite{G-K}.

\begin{figure}
 \begin{center}
  \includegraphics[width=\columnwidth, clip=true]{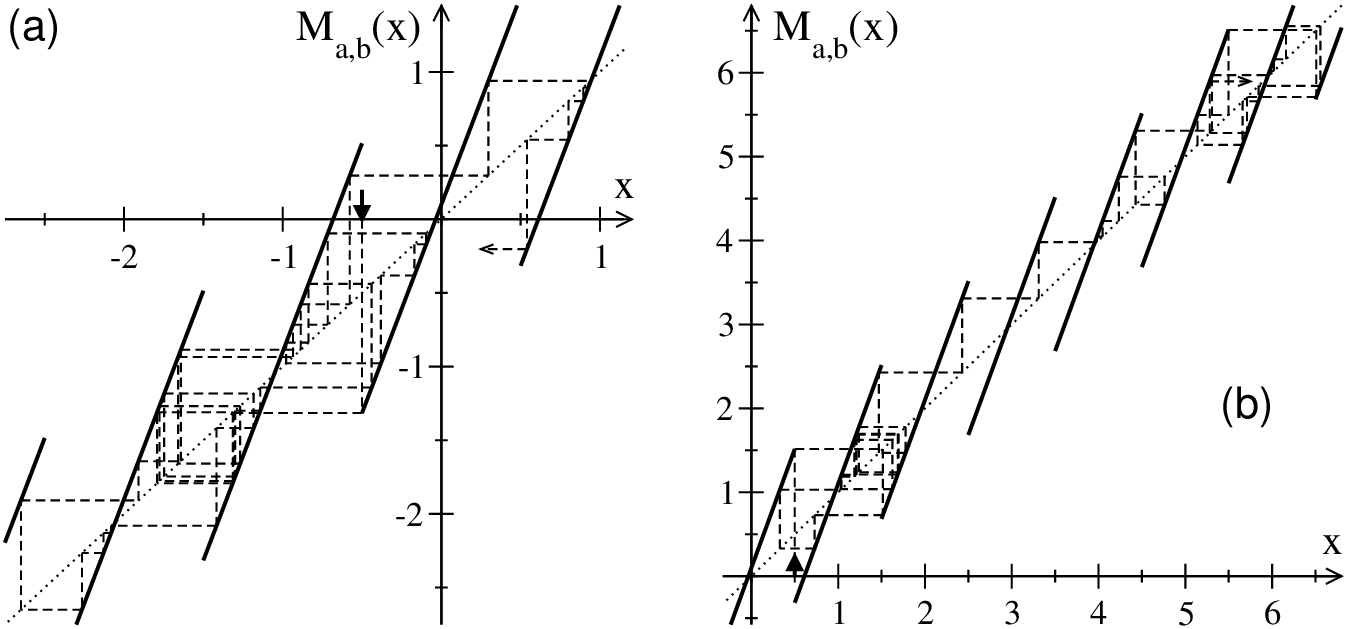}
  \caption{
    \label{FigModel}
    (a) The map $M_{a,b}(x)$ for particular values of the slope $a=2.83$ and the
    bias $b=0.1$ (thick solid lines); the dashed lines represent the first
    33 terms of the trajectory of the point $x_0$  under subsequent
    iterations of $M$ with $x_0 = -\half$ and the fundamental interval $I_0^-$.
    The arrow indicates the initial position $x_0$.\\
    (b) The same as in (a), but for $x_0 = \half$ and the fundamental interval
    $I_0^+$. In both cases the transport coefficients are $J\approx 0.128$ and
    $D\approx0.0543$.

 }
 \end{center}
\end{figure}

The main features of the dynamical system defined by map $M_{a,b}$ are
depicted in figures \ref{FigModel}a and \ref{FigModel}b. They show
$M_{a,b}(x)$ for $a=2.83$ and $b=0.1$ and its action on two initial points
$x_0$. In particular, \fref{FigModel}a presents the trajectory of $x_0=-\half$
using the fundamental interval $[-\half,\half)$, while \fref{FigModel}b shows
the trajectory of $x_0=\half$ with the fundamental interval $(-\half,\half]$.
We chose these two trajectories because, as will be explained, they determine
all transport properties of the system. The trajectories seem to be rather
weakly correlated (which is typical for $b\neq0$) and ``random''. This
randomness is related to the fact that, for nearly all values of $a$ and $b$,
the iteration of the deterministic map $M_{a,b}$ is equivalent to a stochastic
Markov process of a random-walk type \cite{Vollmer03,Nicolis88,Claes93}: the
deterministic trajectory of any point $x_0$ can be regarded as a particular
realisation of a random-walk process, and taking the average over the initial
states $x_0$ is equivalent to taking the averages over the corresponding Gibbs
ensemble. For this reason the process defined by $M_{a,b}$ (or similar maps)
is often called ``deterministic diffusion''. The two basic transport
coefficients, the drift velocity $J$ and diffusion constant $D$, are defined as
\begin{equation}
  J = \lim_{n\to\infty} \frac{\langle x_n\rangle}{n}, \qquad
  D = \lim_{n\to\infty} \frac{\langle x_n^2\rangle - \langle x_n \rangle^2}{2n},
\end{equation}
where $\langle\cdots\rangle$ denotes the average over the uniform ensemble of
initial values $x_0$. Figure~\ref{FigModel} illustrates also the role of using
different fundamental intervals. Had we generated both trajectories using the
same fundamental interval, they would differ only by a simple translation
defined by \eref{eq:defM.1}.

\subsection{Explicit form of the transport coefficients}
We will now briefly summarize Groeneveld and Klages's  \cite{G-K} method of
calculating the transport coefficient for the model defined by the map
$M_{a,b}$. For each  $a>1, b\in \mathbb{R}$, and $\epsilon = \pm$ one can
define two infinite sequences of numbers, $y_r^\epsilon$ and $n_r^\epsilon$,
$r = 0,1,\ldots$, consisting of reals and integers, respectively. Their values
are uniquely determined by demanding that their first terms are given by
\begin{equation}
 y_0^\epsilon = \frac{\epsilon}{2},
 \qquad
 n_0^\epsilon = 0,
\end{equation}
and for each $r > 0$
\begin{equation}
\label{2Reccursion}
n_r^\epsilon + y_r^\epsilon = a y^\epsilon_{r-1} + b,
\end{equation}
with additional conditions
\begin{equation}
  n_r^\epsilon \in \mathbb{Z},
  \quad  \textrm{and} \quad
  y_r^\epsilon \in  I^\epsilon_0.
\end{equation}

Next we define ``N-moments'':
\begin{eqnarray}
\label{Nr}
  N_r^\epsilon &=& -\frac{\epsilon}{2} + \sum_{s=1}^r n_s^\epsilon,\\
\label{Nkl:eps}
  N_{k,l}^\epsilon &=&
     \frac{1}{k!l!}\sum_{r=0}^\infty a^{-r}(N_r^\epsilon)^k r^l,\\
\label{Nkl}
  N_{k,l} &=& N^+_{k,l} - N^-_{k,l},
\end{eqnarray}
where $k,l \ge 0$. The basic transport coefficients, $J$ and $D$ can be now
expressed (\cite{G-K}, cf.\ \cite{Koza99}) as
\begin{eqnarray}
\label{eq:J-D}
  J = \frac{N_{2,0}}{N_{1,1}}, \qquad
  D = \frac{N_{3,0} - N_{2,1}J + N_{1,2} J^2}{N_{1,1}}.
\end{eqnarray}
In the particular case of $b=0$ we have, by symmetry,  $J=0$; hence the
diffusion coefficient takes on a simpler form
\begin{equation}
  \label{defD:b=0}
  D = \frac{N_{3,0}}{N_{1,1}}.
\end{equation}

We will also use two important properties derived in Ref. \cite{G-K}. First
\begin{equation}
 \label{N11>0}
 N_{1,1} > 0,
\end{equation}
so that $D$ and $J$ are well-defined for any  $a\neq0$ and $b$. Second, all
numbers $n_r^\epsilon$ are bounded:
\begin{equation}
 \label{bounded:n}
 |n_r^\epsilon| < \half a + |b|.
\end{equation}

\subsection{Minkowski-Bouligand fractal dimension of the graph of a continuous function}
\label{sub:Method}

Let $E$ be a bounded set in the plane. The $\epsilon$-Minkowski sausage of $E$,
denoted by $E(\epsilon)$, is the set of all the points whose distance to $E$ is
less then $\epsilon$. Let $\mathcal A(E(\epsilon))$ be the area of
$E(\epsilon)$. Then the Minkowski-Bouligand fractal dimension is defined as
\cite{Tricot}
\begin{equation}
  \Delta(E) = \lim_{\epsilon\to 0} \left( 2 - \frac{\log \mathcal A (E(\epsilon))}{E(\epsilon)}
  \right),
\end{equation}
provided that this limit exists.  In this case it is equivalent to the
box-counting dimension \cite{Tricot}. A general relation between the
Minkowski-Bouligand dimension $\Delta$ and the Hausdorff dimension $\Delta_H$
reads $\Delta \ge \Delta_H$. It is conjectured that for all strictly
self-similar fractals $\Delta = \Delta_H$.

It has been recently proved  \cite{G-K} that the transport coefficients
corresponding to the map $M_{a,b}$ are continuous (but not differentiable)
functions of the slope $a$ even though $N_{j,k}$ are generally not continuous
in $a$. Therefore we can calculate $\Delta$ for the graphs of $D(a)$ or $J(a)$
using the fact that the Minkowski-Bouligand dimension of a continuous real
function $f(t)$ defined on an interval $[t_0,t_1]$ can be evaluated through
analysis of its H\"older exponents \cite{Tricot} and $\tau$-oscillations
\cite{Tricot}. The latter are defined as
\begin{equation}
\label{def:osc}
  \textrm {osc}_\tau(f;t) =
   \sup_{|t-t'|\le \tau}f(t') - \inf_{|t-t'|\le \tau}f(t').
\end{equation}
The H\"older exponents are related to the oscillations as follows. If
there exist constants $c>0$ and $0 < H \le 1$ such that for all $\tau$
\begin{equation}
 \label{def:Holderian}
  \textrm{osc}_\tau(f;t) \le c\tau^H
\end{equation}
then $f$ is called a Holderian of exponent $H$ at $t$. If constants $c$ and
$H$ are independent of $t$ then the fractal dimension $\Delta$ of $f$ satisfies
\begin{equation}
 \label{DeltaLE}
  \Delta \le 2-H.
\end{equation}
Similarly, if there exist constants $c>0$ and $H$ such that
for all $\tau$
\begin{equation}
 \label{def:AntiHolderian}
  \textrm{osc}_\tau(f;t) \ge c\tau^H
\end{equation}
then $f$ is called an anti-Holderian of exponent $H$ at $t$; if $c$ and $H$
are independent of $t$ then
\begin{equation}
 \label{DeltaGE}
  \Delta \ge 2-H.
\end{equation}
Relations (\ref{DeltaLE}) and (\ref{DeltaGE}) constitute a convenient means of
calculating the  Minkowski-Bouligand dimension $\Delta$. Note that although they
are supposed to hold ``for all $\tau$'', in practice it suffices to prove their
validity only in the limit of $\tau  \to 0$.

\subsection{Conjecture about the logarithmic correction to the fractal dimension}
\label{sub:Conjecture}

In \cite{Koza04} we proposed the following conjecture: for the map $M_{a,b}$
in the limit of $\tau \to 0$
\begin{equation}
 \label{conjecture}
  \frac{\textrm{osc}_\tau(D;a)}{\tau} \simeq
  {\mathcal C}(a)\left[-\log(\tau)\right]^{\gamma(a)}, \quad \mathrm{with}\; \mathcal C(a),\gamma(a) > 0.
\end{equation}
Together with  (\ref{def:osc}) --
(\ref{DeltaGE}) this implies that the fractal dimension of the
graph of the diffusion coefficient $D$ as a function of the control parameter
$a$ is equal 1, but the convergence to this limiting value is logarithmically
slow, with the exponent $\gamma$ controlling the convergence rate.

To justify this statement note that since $\lim_{\tau\to 0} (-\log
\tau)^\gamma/\tau^{H-1} = 0 $ for any $\gamma \ge 0$ and  $H<1$, conjecture
(\ref{conjecture}) implies that $D(a)$ is  Holderian of any $H<1$. Using
(\ref{DeltaLE}) we find that the fractal  dimension of its graph is bounded
from above by any number of the form $2 - H$, with $H<1$. Now we can take the
limit of $H \uparrow 1$ to see that $\Delta$ cannot exceed 1. Similarly
(\ref{conjecture}) implies that $D(a)$ is  anti-Holderian of exponent $H=1$,
and so $\Delta \ge 1$. Hence $\Delta = 1$.

In the above analysis we have tacitly assumed that the coefficient $\mathcal
C(a)$ in (\ref{conjecture}) can be bounded from below and above by  positive
constants independent of $a$. We will address this problem in section
\ref{subsub:plot-of-c}.

Since $D$ is a continuous, nowhere differentiable function of $a$ \cite{G-K},
the fractal dimension $\Delta$ must be $\ge 1$  and exponent $\gamma$ must be
strictly positive. This reasoning constitutes an alternative derivation of the
lower bound for $\Delta$ and explains why $\gamma(a) > 0$. Our numerical
calculations (performed for the bias $b=0$) indicated that exponent $\gamma$
depends on the slope $a$ and is equal either to 1 or 2, depending on
periodicity properties of the sequences $y_k^\pm$ \cite{Koza04}. In
particular, we conjectured that $\gamma=2$ for the slopes $a$ generating
disjoint sequences $y_k^+$ and $y_k^-$, and $\gamma = 1$ otherwise.

Note that even though $\Delta= 1$, the graph of $D(a)$ is not an ordinary,
smooth curve of  dimension 1. Actually, owing to the logarithmic term in
(\ref{conjecture}), this graph  is nowhere rectifiable (i.e., any of its arcs
is ``of infinite length''), which is typical of fractals \cite{Tricot}.
Therefore the graph of $D(a)$ locally resembles a well-known family of Takagi
functions \cite{KlagesHab,KlagesPHD,DorfmanBook,Tricot}, for their graphs are
also nonrectifiable, their  dimension is $\Delta = 1$, and they satisfy
(\ref{conjecture}) with $\gamma(a) = 1$ for all $a$. Takagi functions appear
naturally in many contexts related to deterministic diffusion
\cite{KlagesHab,Gilbert01,KlagesPHD,DorfmanBook}.

Since hypothesis (\ref{conjecture}) opens a convenient way to explore fractal
properties of graphs of the diffusion coefficient, its detailed analysis,
carried out with both analytical and numerical methods, is one of the main
goals of our present paper.

\section{Fractal dimension of the diffusion coefficient for $b=0$}
\label{sec:case.b=0}
In the case of zero bias ($b=0$) the diffusion coefficient $D$ vanishes for
the slopes $a \le2$. Therefore we shall consider only the nontrivial case of
\begin{equation}
\label{a>2}
  a > 2.
\end{equation}

The basic properties of the diffusion coefficient are encoded in the sequences
$(n_k^\epsilon)$, introduced in Sec.~\ref{sub:model}. Before we will be able
to tackle the problem of evaluating the fractal dimension of $D(a)$, we need to
derive several useful properties of these sequences.

\subsection{Basic properties of $n_k$ and $y_k$ for $b=0$}
\label{sub:BasicProperties}

\subsubsection{Symmetry.}
For $b=0$ the sequences $n_k^\pm$ and $y_k^\pm$ satisfy
\begin{equation}
  \label{nk=-nk}
  n_k^+ = -n_k^-,\qquad y_k^+ = -y_k^- ,
\end{equation}
so it will suffice to concentrate on the sequences $n_k^+$ and $y_k^+$; for
simplicity we will denote their terms as $n_k$, $y_k$, respectively. These
sequences may be either periodic or not. Periodic sequences $n_k$ (and $y_k$)
correspond to the so called Markov partitions of the interval $I_0^+$ (or
$I_0^-$) \cite{KlagesPRE99,KlagesPHD,Vollmer03,Nicolis88}. Parameters $a$ for which
sequences $n_k$ and $y_k$ are periodic will be called ``Markov slopes''. They
form a dense set on $(1,\infty)$ \cite{G-K}.

\subsubsection{Discontinuity in $a$.}
\begin{figure}
  \includegraphics[width=0.975\columnwidth, clip=true]{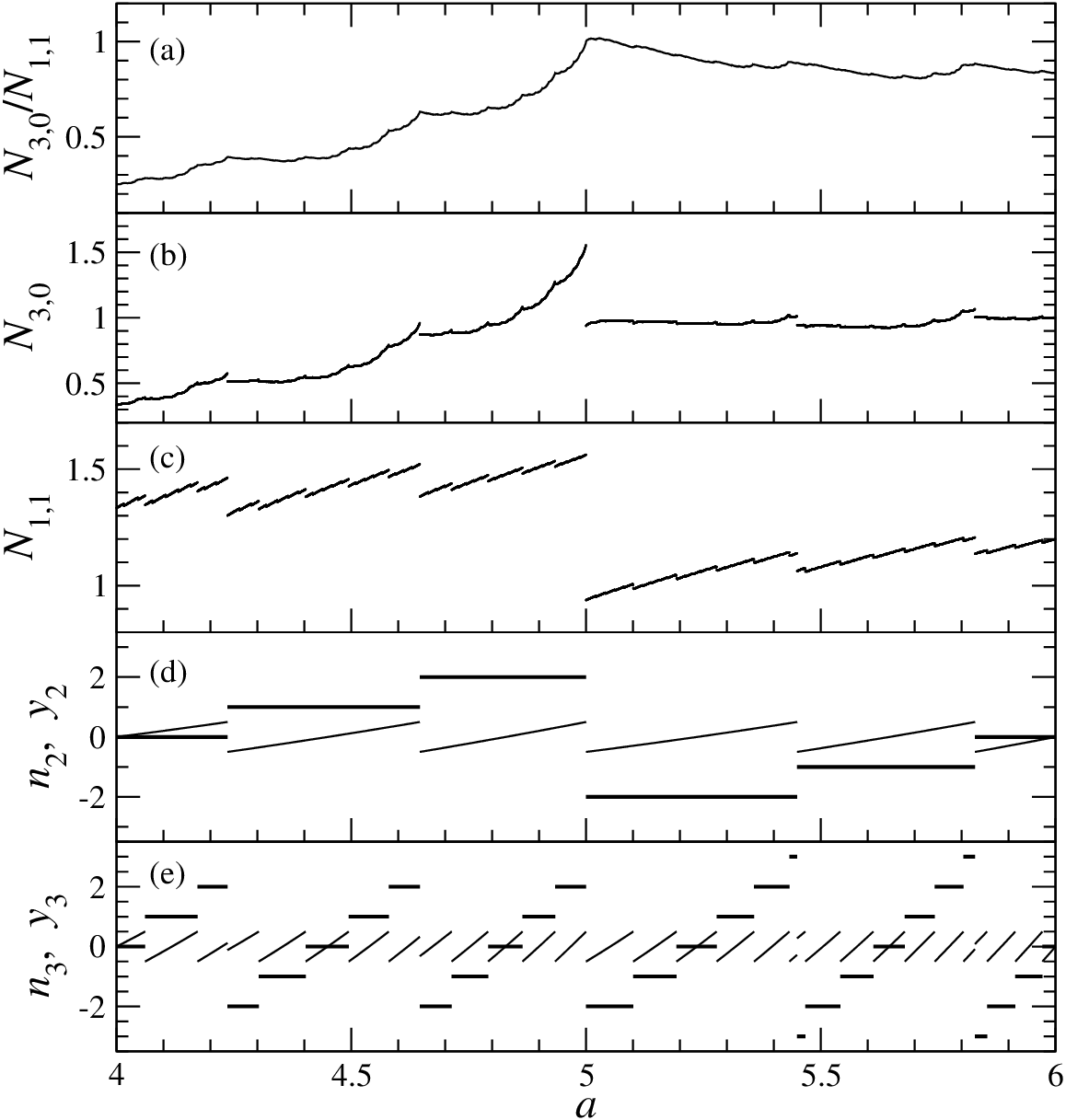}
  \caption{
    \label{Fig1} Graphs of the basic quantities used in the mathematical
    derivation as functions of the slope $a$, calculated for the bias $b=0$.
 (a) The diffusion coefficient $D = N_{3,0}/N_{1,1}$; (b) $N_{30}$; (c)
 $N_{11}$; (d) $n_2$ (horizontal, thick lines) and $y_2$ (thinner curves);
 (e)~$n_3$ (horizontal, thick lines) and $y_3$ (thinner, almost linear
 curves). Except for $D$, all of them are discontinuous.  $N_{jk}$ are
 continuous on a dense set $\bar\Xi$ and discontinuous on a dense set $\Xi$
 which is the union of all discontinuity points of $n_1(a), n_2(a),
 n_3(a)\ldots$ (see text for details). }
\end{figure}

The numbers $n_k$, as well as $y_k$, can be considered as functions of $a$.
Equation \eref{2Reccursion} implies that all functions $n_k(a)$ and $y_k(a)$,
$k=1,2,\ldots,$ are discontinuous in $a$, which has a profound impact on the
functional dependence of the transport coefficients $J$ and $D$ on $a$. This
is illustrated in \fref{Fig1}. \Fref{Fig1}a presents the quantity of our
primary interest: the diffusion coefficient $D$ as a function of the slope $a$
for $4\le a \le 6$ and for the vanishing bias $b=0$. Although highly
irregular, this function is continuous \cite{G-K}. One might expect that so
are $N_{3,0}(a)$ and $N_{1,1}(a)$, because their quotient equals to $D$.
However, as depicted in figures~\ref{Fig1}b and~\ref{Fig1}c, actually they are
discontinuous. From (\ref{Nr}) -- (\ref{Nkl}) we conclude that this must be
related to discontinuity of functions $n_k^\pm(a)$. Two of them, $n_2^+(a)$ and
$n_3^+(a)$, are shown in figures~\ref{Fig1}d and \ref{Fig1}e, respectively (as
horizontal line segments). For completeness, in the same figures are also
depicted the graphs of $y_2^+(a)$ and $y_3^+(a)$. As expected, there is a
clear correlation between discontinuities of $n_k(a)$, $y_k(a)$ and
discontinuities of $N_{3,0}(a)$, $N_{1,1}(a)$. Moreover, there is also a
correlation between these discontinuities and ``irregularities'' of $D(a)$.

The union of all discontinuity points of $n_k(a)$ for all $k$ will be denoted
by $\Xi$ and we will introduce $\bar\Xi$ as the complement of $\Xi$ on the
interval $(1,\infty)$. Importance of these two sets is related to the fact
that functions $N_{k,l}(a)$ are continuous on $\bar\Xi$ and discontinuous on
$\Xi$. Moreover, functions $n_k(a)$ and $y_k(a)$ are also continuous on
$\bar\Xi$ (however, in contrast to $N_{k,l}(a)$, they can be also continuous
on some points of $\Xi$). Both $\Xi$ and $\bar\Xi$ are dense on $(1,\infty)$
\cite{G-K}.

\subsubsection{Approximate linearity of $y_k(a)$.}
\label{subsubsec:right}
Figures \ref{Fig1}d and \ref{Fig1}e suggest that $y_2(a)$ and
$y_3(a)$ consist of almost linear segments even though in reality they are
polynomials of degree $2$ and $3$, respectively. Moreover, the density of
discontinuity points of $n_3$ is considerably greater than that of $n_2$.
These two important observations can be justified in a more formal way. First
of all note that \eref{2Reccursion} implies
\begin{equation}
  \label{yk'}
    y_k' = y_{k-1} + a y_{k-1}'
         = \sum_{j=0}^{k-1} y_{k-1-j} a^j,
\end{equation}
where we used a short-hand notation $y_k' = \rmd y_k(a)/\rmd a$. Therefore,
since $|y_k(a)| \le \half$ and $y_0 = \half$, the derivative $\rmd y_k(a)/\rmd
a$ is bounded
\begin{equation}
  \label{Bound}
       \frac{a^{k-1}(a-2)+1}{2(a-1)} \le y_k' \le  \frac{a^k-1}{2(a-1)}.
\end{equation}
Taking into account \eref{a>2} we conclude that for large $k$
\begin{equation}
\label{y_k_prim}
    y_k' \sim a^{k-1}.
\end{equation}
Since the range of each $y_k$ is limited to $[-\half, \half]$, the distance
between two consecutive discontinuity points of $y_k(a)$, denoted as $\delta_k
(a)$, is of order $\lesssim a^{1-k}$, i.e.\ decreases rapidly with $k$.
Similar reasoning leads to the conclusion that the second derivative of
$y_k(a)$ is of order $\lesssim 2(k-1)a^{k-1}$. Thus the maximal error
resulting from truncating the Taylor series of $y_k(a)$ after the linear term
for the displacement $\delta_k (a) \sim a^{1-k}$ is of order $(k-1)/a^{k-1}$
and vanishes quickly as $k$ goes to infinity. Therefore functions $y_k(a)$ can
actually be approximated to a high accuracy by piece-wise linear segments.

\subsubsection{The limits of $a$ going to $a_0$ from below and from above.}
Although functions $N_{k,l}(a)$ are discontinuous on $\Xi$, they are left
continuous. Moreover, it turns out that for any $a_0\in (1,\infty)$ the limit
of $N_{k,l}(a)$ for $a$ going to $a_0$ from above is also well-defined. We
will denote it by $\tilde N_{k,l}(a_0)$. Its value can be related to
$N_{k,l}(a_0)$ as follows. For $a_0\in \bar\Xi$, by definition of
$\tilde\Xi$,  $\tilde N_{k,l}(a_0) = N_{k,l}(a_0)$. On the other hand, for
$a_0\in\Xi$ the sequence $n_0(a_0),n_1(a_0),\ldots$ is strictly periodic.
Denoting the period length by $L$ we thus have $n_k(a_0) = n_{k+L}(a_0)$ for
all $k\ge0$. For $a$ tending to $a_0$ from above the sequence $n_{k}(a)$ will
converge to a new sequence $\tilde n_k(a_0)$
\begin{equation}
  \label{def:tilde_n}
   \tilde n_k(a_0) =
     \cases{
       n_k(a_0),   & $k < L$ \\
       n_k(a_0)+1, & $k=L$ \\
       -n_k(a_0),  & $k > L$
     }
\end{equation}
Now $\tilde N_{k,l}(a_0) = \lim_{a\to a_0+} N_{k,l}(a_0)$ can be determined
upon substituting $\tilde n_k(a_0)$ for $n_k$ in \eref{Nr} -- \eref{Nkl}.
Functions $\tilde N_{k,l}(a)$ are right continuous. As argued in \cite{G-K},
all transport coefficients of the system defined by the map $M_{a,b}$ are
continuous in $a$. Consequently, functions $\tilde N_{k,l}(a_0)$ can replace
$N_{k,l}(a)$ in \eref{eq:J-D} to calculate the transport coefficients of the
system. It is also interesting to note that even though the functions
$N_{k,l}(a)$ and $\tilde N_{k,l}(a)$ are different, their graphs must be the
same. For this reason figure \ref{Fig1}b actually shows \emph{both}
$N_{3,0}(a)$ and $\tilde N_{3,0}(a)$. Similar conclusion pertains also to
\fref{Fig1}c.

\subsection{Oscillations at elements of $\bar\Xi$}
\label{sub:bar.Xi}

Let $a_0$ be an arbitrary element of $\bar\Xi$. This means that all functions
$y_k(a)$ are continuous at $a_0$, and so for each integer $k$ there exists
$\varepsilon_0 > 0$ such that  $y_0, y_1,\ldots,y_k$ and $n_0, n_1,\ldots,n_k$
are continuous and differentiable functions of $a$ on the interval $(a_0
-\varepsilon_0, a_0 + \varepsilon_0)$. Therefore for any $a \in (a_0
-\varepsilon_0, a_0 + \varepsilon_0)$ all functions $n_j(a)$ up to $j=k$ are
independent of $a$. Let $0 < \tau < \varepsilon_0$. Since $y_k(a)$ is
continuous on $(a_0 -\varepsilon_0, a_0 + \varepsilon_0)$, we can write down
its Taylor expansion about $a_0$ and, following our discussion in
\sref{subsubsec:right}, truncate it after the linear term, obtaining
$y_k(a_0+\tau) \approx y_k(a_0) + \tau y_k'(a_0)$.

Let $\mu(\tau; a)$ be the maximum integer $j$ such that $y_{j}(a)$ and
$n_{j}(a)$ are continuous on $(a -\tau, a + \tau)$.  This means that for any
$a_0\in\bar\Xi$ and a fixed $\mu$ numbers $\mu$ and $\tau$ are related through
$y_\mu(a_0) + \tau y_\mu'(a_0) \approx \half$ or $y_\mu(a_0) - \tau
y_\mu'(a_0) \approx -\half$ (whichever gives  smaller $\tau$). This, together
with \eref{y_k_prim}, leads to the following  implicit equation for $\mu(\tau;
a)$:
\begin{equation}
 \label{mu:general}
  \mu(\tau; a) \approx \log_{a}\frac{\half-|y_{\mu(\tau; a)}|}{\tau}.
\end{equation}

There are two cases: either the sequence $y_0(a_0), y_1(a_0),\ldots$ is
periodic (i.e.\ the system has a Markov partition) or not. In the former case
the numerator in \eref{mu:general} is bounded from above and below by
\emph{positive} numbers; in the limit of $\tau\to 0$ we thus have
\begin{equation}
  \label{m-limit}
  \mu(\tau; a) \approx \log_a\tau^{-1} + \mathrm{const}
    \qquad (\mbox{for } a \in \bar\Xi).
\end{equation}
Clearly $\mu(\tau;a) \to \infty$ as $\tau\to 0$. In other words, in the limit
$\tau\to0$ (which is essential for the analysis of H\"older exponents), the
number of functions in the sequences $y_k(a)$ and $n_k(a)$ that are
continuous on $(a-\tau, a+ \tau)$ goes to infinity.

The case where the sequence $y_0(a_0), y_1(a_0),\ldots$ is not periodic is far
more complicated. The main problem comes from the fact that now the numerator
in \eref{mu:general} can get arbitrarily close to zero, which may invalidate
\eref{m-limit}. In the present approach we restrict ourselves to the cases
where the system has a Markov partition and so the sequence  $y_0(a_0),
y_1(a_0),\ldots$ is periodic.

As expressed by (\ref{defD:b=0}), the diffusion coefficient is the quotient
of two quantities, $N_{3,0}$ and $N_{1,1}$. However, by a straightforward
calculation we can verify that for any  $\tau > 0$ and any functions $f(t)$
and $g(t)$ bounded on $[t-\tau, t+\tau]$ (no matter if continuous or not, but
the lower bound for $|g|$ must be $\neq 0$)
\begin{eqnarray}
 \label{osc_upper_bound}
       \textrm{osc}_\tau(f/g)
       &=&
          \textstyle \max_\tau(f/g) - \min_\tau(f/g)        \nonumber \\
       &\le&
          \frac{\textrm{osc}_\tau(f) \max_\tau(g) +
                \textrm{osc}_\tau(g) \min_\tau(f)}{
               \min_\tau(g) \max_\tau(g)}             \nonumber \\
       &\le&
            c_1\textrm{osc}_{\tau}(f)  + c_2\textrm{osc}_{\tau}(g)
\end{eqnarray}
where we used a short-hand notation $\max_\tau(f) \equiv \sup_{|t-t'| \le
\tau} f(t')$ and $\min_\tau(f) \equiv \inf_{|t-t'| \le \tau} f(t')$, while
$c_1$ and $c_2$ are some nonnegative parameters independent of $\tau$.

Thanks to (\ref{N11>0}) we can apply (\ref{osc_upper_bound}) to $D(a)$,
substituting $N_{3,0}$ for $f$, $N_{1,1}$ for $g$, and $a$ for $t$. The problem
of finding the $\tau$-oscillations of the diffusion coefficient $D$ can be
thus reduced to that of finding $\textrm{osc}_\tau(N_{3,0};a)$ and
$\textrm{osc}_\tau(N_{1,1};a)$.

Equation (\ref{Nkl:eps}) implies that for arbitrary integers $k,l$ and $a =
a_0 \pm \tau$
\begin{equation}
\label{N_kl(a):0}
  \fl
  N_{k,l}(a_0) - N_{k,l}(a) \le \frac{2}{k!l!}\sum_{r=\mu(\tau; a_0)}^\infty
  \left\{a_0^{-r}[N_r^+(a_0)]^k - a^{-r}[N_r^+(a)]^k \right\}
  r^l,
\end{equation}
with $\mu$ related to $\tau$ through equation (\ref{m-limit}).  Here we have
employed the fact that, by definition of $\mu$, $n_r(a) = n_r(a_0)$ and hence
$N_{r}^+(a) = N_{r}^+(a_0)$ for all $r<\mu(\tau;a)$. Using (\ref{bounded:n})
we conclude that
\begin{equation}
\label{eq:c1}
|N_r^+| \le c^*_1r + c^*_2,
\end{equation}
with $c_1^*\ge0, c_2^* > 0$ being some constants. It will be convenient to
assume that $c^*_1$ denotes the smallest possible value satisfying
\eref{eq:c1}, $c^*_1 \equiv \limsup_{r\to\infty} |N_r^+(a_0)|/r$.

Thanks to \eref{eq:c1} an upper (as well as lower) bound for $N_{k,l}(a_0) -
N_{k,l}(a)$ can be found by substituting arithmetic sequences for $N_r^+(a_0)$
and $N_r^+(a)$ in \eref{N_kl(a):0}. Their common differences may be different,
but the terms for $r=\mu$ must be the same or differ by a small number less
than $\half a + |b|$, cf \eref{bounded:n}. This leads to
\begin{eqnarray}
 \label{|N_kl(a)|:1}
  |N_{k,l}(a) - N_{k,l}(a_0)|   \lesssim a^{-\mu} \mu^{k+l-1},
   &&\qquad\textrm{for } c^*_1>0, \\
  |N_{k,l}(a) - N_{k,l}(a_0)|   \lesssim a^{-\mu} \mu^{l},
   &&\qquad\textrm{for } c^*_1=0.
\end{eqnarray}
This important result can be also obtained by noticing that the major contribution
to the sum in \eref{N_kl(a):0} comes from its first few terms.
Hence, using \eref{m-limit},
\begin{eqnarray}
  |N_{k,l}(a_0\pm\tau) - N_{k,l}(a_0)|
       \lesssim
       \tau \ln^{k + l -1}(\tau^{-1}), &\qquad& c_1^* > 0,\\
  |N_{k,l}(a_0\pm\tau) - N_{k,l}(a_0)|
       \lesssim
       \tau \ln^{l}(\tau^{-1}), &\qquad& c_1^* = 0,
\end{eqnarray}
which implies
\begin{eqnarray}
  \label{eq:gen.osc.bound.1}
  \textrm{osc}_\tau(N_{k,l};a) \lesssim \tau \ln^{k + l -1 }(\tau^{-1}), &\qquad& c_1^*>0,\\
  \label{eq:gen.osc.bound.2}
  \textrm{osc}_\tau(N_{k,l};a) \lesssim \tau \ln^{l}(\tau^{-1}), &\qquad& c_1^*=0.
\end{eqnarray}
Consequently, by \eref{osc_upper_bound} and \eref{defD:b=0},
\begin{eqnarray}
\label{c1:>0}
 \textrm{osc}_\tau(D;a) \lesssim  \tau \ln^2 (\tau^{-1}),&& \qquad c_1^*>0,\\
\label{c1:=0}
 \textrm{osc}_\tau(D;a) \lesssim  \tau \ln (\tau^{-1}),&& \qquad c_1^*=0.
\end{eqnarray}
for all $a\in\bar\Xi$.


Equations (\ref{c1:>0}) and (\ref{c1:=0}) give the upper bound for the exponent $\gamma$
controlling the
logarithmic correction to the local fractal dimension. As we can see, this upper bound depends
 on
$c^*_1$. This constant has a simple physical interpretation. If the
trajectories of points $x_0=\pm\half$ under $M_{a,b}$ (like those depicted in
\fref{FigModel}) remain confined inside a neighborhood of the origin, we can
take $c^*_1 =0$, which implies $0 < \gamma\le1$. Otherwise $c^*_1>0$ and $0 <
\gamma \le 2$.

\subsection{Oscillations at elements of $\Xi$}
\label{sub:Xi}

The argumentation presented in section~\ref{sub:bar.Xi}  might seem of little
use for elements of the set $\Xi$, at which functions $N_{k,l}(a)$ are
discontinuous. However functions $N_{k,l}(a)$ are left continuous and have a
well-defined limit from the right, $\tilde N_{k,l}(a)$. Functions $\tilde
N_{k,l}(a)$ are right continuous and, as argued in
 \sref{subsubsec:right}, can replace $N_{k,l}(a)$ in calculations of
the transport coefficients. This suggests a method of circumventing the
problems related to discontinuity. The oscillation of the diffusion
coefficient $D$ on the interval $[a_0-\tau,a_0+\tau]$ with $a_0\in\Xi$ can be
calculated as the maximum of two values: the oscillation of
$(N_{3,0}/N_{1,1})(a)$ on $[a-\tau,a]$ and the oscillation of $(\tilde
N_{3,0}/\tilde N_{1,1})(a)$ on $[a,a+\tau]$. In each case we can now apply the
reasoning of section~\ref{sub:bar.Xi} (its most delicate point, validity of
approximation \eref{m-limit}, can be justified rigorously because the sequence
$n_k(a)$ is periodic for all $a\in\Xi$). This implies that \eref{c1:>0} and
\eref{c1:=0} are valid also for all $a\in \Xi$, which suggests that the fractal
dimension of the diffusion coefficient is 1 for all Markov slopes $a \in
(1,\infty)$.

\section{The case $b\neq0$}
\label{sec:b.neq.0}
Besides the fractal properties of the diffusion coefficient $D$, in the
asymmetric case $b\neq0$ we can investigate also those of the drift
$J$, which for $b\neq0$ is also a highly irregular function of the
control parameters \cite{G-K}. It turns out that the methods developed in
\sref{sec:case.b=0} can be readily extended to the case of nonvanishing bias
$b$ and to the analysis of arbitrary transport coefficients. Therefore we will
only discuss the major  differences and difficulties appearing when applying
these methods to the more general case $b\neq0$.

First, although derivation of the basic relation \eref{mu:general} remains
practically unchanged, we see that restricting our calculations to the case
$1<a\le 2$ is fully acceptable only for $b=0$. This is related to the fact
that for $b\neq 0$ the transport coefficients can assume nontrivial values
also for $a\le2$ \cite{G-K}, so this parameter region also deserves attention.
The assumption $a>2$ greatly simplifies the proof of \eref{y_k_prim}
(cf.~\eref{Bound}); without it we have to take into account correlations
between consecutive terms of the sequences $y_k^\pm$. Our numerical analysis
shows that \eref{Bound}, and hence \eref{mu:general}, is valid also for
$1<a\le2$ (at least in the limit of $k\to\infty$), but we failed to find a
rigorous proof of it.

Second, the sequences $n_k^+$ and $n_k^-$ are no longer related to
each other by \eref{nk=-nk} and in practice must be considered as
independent of each other. This, fortunately, is not a serious
difficulty -- actually relation \eref{nk=-nk} was used mainly to
simplify notation.

Third, just as in the case $b=0$, we can justify transition from
\eref{mu:general} to \eref{m-limit} only in the case where both $n_k^+$ and
$n_k^-$ are periodic, i.e.\ when the system has a Markov partition.

Fourth, we still need an analytical argument justifying our claim that
the graph of $(D;a)$ is \emph{uniformly} anti-Holderian for all $b$.

Although in \sref{sec:case.b=0} we focused on the fractal properties of the
diffusion coefficient $D$, our analysis was as general as possible. In
particular inequalities \eref{eq:gen.osc.bound.1} and
\eref{eq:gen.osc.bound.2} were derived for  functions $N_{k,l}(a)$ with
arbitrary $k,l$. Since all transport coefficients, including the drift
velocity $J$, can be expressed as functions of the quotients $N_{k,l}/N_{1,1}$
\cite{G-K}, the methods  of \sref{sec:case.b=0} can be applied for arbitrary
transport coefficients.

Consequently, we can follow the ideas of \sref{sec:case.b=0} to argue that the
local fractal dimension of the graphs of transport coefficients as functions
of $a$ is 1 for the control parameters $a$ and $b\in \mathbb{R}$ such that
$a>2$ and the system has a  Markov partition. Based on our numerical
calculations \cite{Koza04} and heuristic arguments we conjecture the same
fractal behaviour for all other pairs $(a,b)$.

\section{Numerical results}
\label{sec:Numerical}

\subsection{The role of $c^*_1$}
\label{subsub:c}
Equations \eref{c1:>0} and \eref{c1:=0} imply that the exponent $\gamma$
controlling the logarithmic correction is bounded from above by 2 for $c^*_1
>0 $ and by 1 for $c^*_1 = 0$.
On the other hand, however, in  \cite{Koza04} we conjectured that the value of
$\gamma$ depends on properties of the periods of sequences $y_k^+$ and $y_k^-$
(i.e., whether they consist of the same or different terms). A question arises
whether  the two approaches lead to the same conclusions for $\gamma$ and if
not, what does its value really depend on?

\begin{figure}
\begin{center}
  \includegraphics[width=0.45\columnwidth, clip=true]{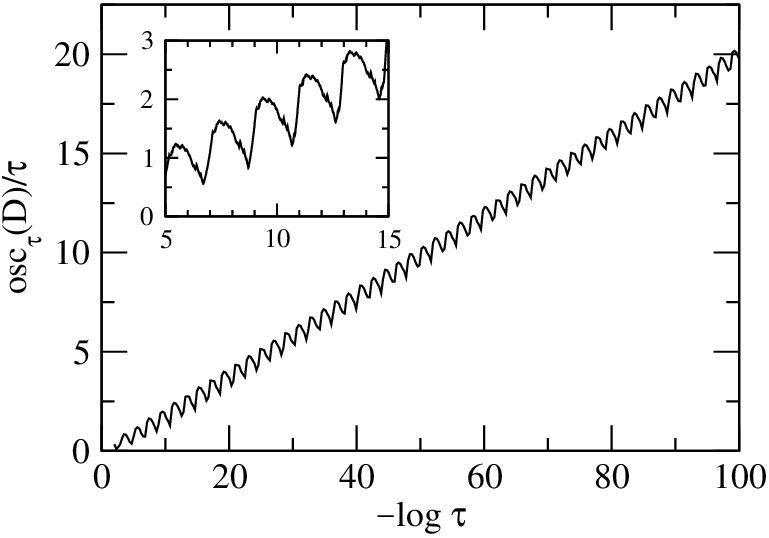}
  \caption{
    \label{Fig3}
    Oscillations of the diffusion coefficient $D$ for $b=0$ and $a$
    equal to the largest root of $a^5 - 2a^4 - 2a^3 + 2a^2 + 1$, i.e.\ $a \approx 2.455$,
    rescaled in accordance with (\protect\ref{conjecture}) and
    calculated on intervals $(a-\tau,a+\tau)$ for $10^{-100} \le \tau \le
    10^{-2}$. The inset presents a blowup of the results and reflects
    periodicity of the curve. }
\end{center}
\end{figure}

To answer it we will study a particular case of the bias $b=0$ and the slope
$a$ equal to the largest root of $a^5 - 2a^4 - 2a^3 + 2a^2 + 1$, i.e.\
$a\approx 2.455$. The trajectory of $x_0^+=\half$ is  $x_1^+\approx 1.23,
x_2^+\approx 1.56, x_3^+\approx 0.92, x_4^+\approx 0.80, x_5^+ =
x_0^+,\ldots$, and so the sequence $x^+_m$ is periodic, $x_{5+m}^+ = x_m^+$.
The trajectory of $x_0^-$ is given by $x_m^- = -x_m^+$, and hence is periodic,
too. Since all terms of the sequence $x_k^-$ are different from those of
$x_k^+$, according to \cite{Koza04} the value of $\gamma$ should be 2. This is
in conflict with \eref{c1:=0}, as the trajectories of $x_0=\pm\half$ are
bounded, hence $c^*_1=0$ and, following \eref{c1:=0}, $\gamma \le 1$. We
checked this numerically using the arbitrary precision arithmetic library GMP
\cite{GMP}. Our results are presented in \fref{Fig3}. It clearly suggests that
$\gamma=1$. This implies that our earlier hypothesis about the logarithmic
correction exponent $\gamma$ was incorrect. Our present, deeper insight into
mathematical nuances of the problem leads to the conclusion that the main
factor determining the value of $\gamma$ (at least for systems with periodic
trajectories of $x_0=\pm\half$) is whether the constant $c^*_1$ vanishes or
not.

Figure \ref{Fig3}, and especially the inset in it, indicates also that in the
case of the map $M_{a,b}$ oscillations of the diffusion coefficient really
deserve their name. Apparently the graph of $\textrm{osc}_\tau(D)/\tau$ as a
function of $\log\tau^{-1}$ is a composition of a linear and periodic,
``oscillating'' function. The period length $\lambda$ of these ``oscillations
of oscillations'' can be estimated using \eref{m-limit},
\begin{equation}
 \label{def:lambda}
 \lambda =  a^L,
\end{equation}
where $L$ is the period of $y_k^\pm$; in our case $L=5$, $\lambda = a^5\approx
89$, $\log\lambda\approx 1.95$, in excellent agreement with the data in
\fref{Fig3}.

\subsection{The uniformly H\"older and anti-H\"older conditions for $D(a)$}
\label{subsub:plot-of-c}

As we explained in section \ref{sub:Conjecture}, analysis of H\"older
exponents can be useful in exploration of fractal properties of a curve if
this curve is uniformly Holderian (or anti-Holderian), i.e.\ if the
coefficient $\mathcal C(a)$ in (\ref{conjecture}) can be bounded from below
(or above) by a constant number. We decided to check these properties
numerically for the vanishing bias $b$, as in this particular case all Markov
slopes $a$ are algebraic numbers. They can be thus generated in a systematic
way by first generating polynomials with integer coefficients and then
identifying Markov slopes with the largest roots of the polynomials. Moreover,
the coefficients of each such polynomial determine the numbers $n_r^\epsilon$,
hence whether the constant $c_1^*$ vanishes or not, and thus whether the
exponent $\gamma(a)$ in our conjecture (\ref{conjecture}) is expected to equal
1 or 2, respectively.

\begin{figure}
\begin{center}
  \includegraphics[width=\columnwidth, clip=true]{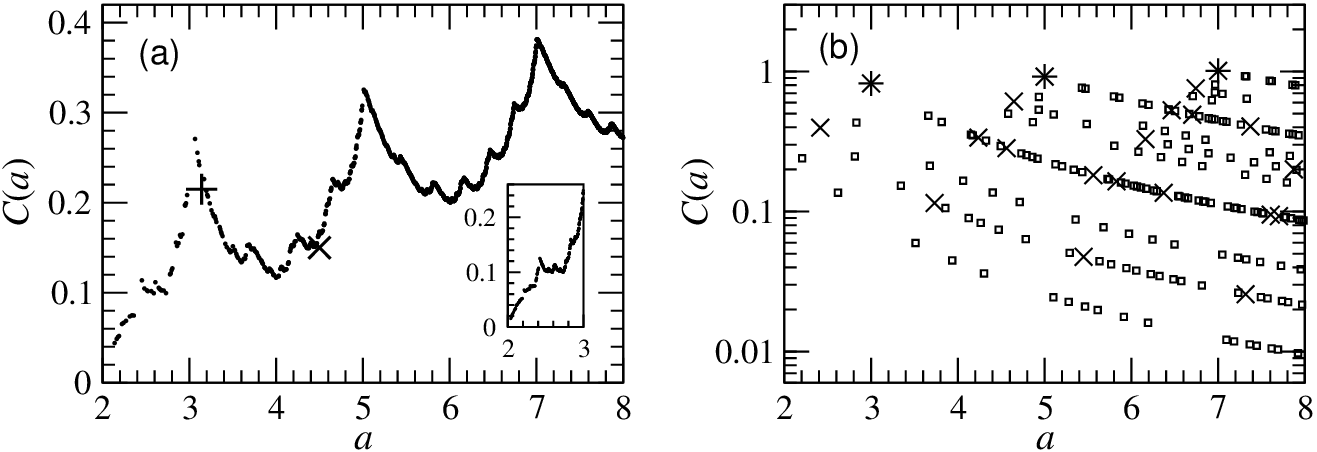}
  \caption{
    \label{Fig4}
(a) The coefficient $\mathcal C(a)$ for all Markov slopes $a\in(2;8]$ that
correspond to $\gamma = 1$ and are algebraic numbers of degree $d \le 4$ (2054
data points). The inset presents the same quantity calculated for $a\in(2;3]$
and $d \le 6$ (439 data points). The plus ($+$) and cross ($\times$) signs mark
the estimated values of $\mathcal C(a)$ for non-Markov slopes $a = \pi$ and
$a=4.5$, respectively. (b) The same as in (a), but for $\gamma = 2$ and $d =1$
(stars), $d=2$ (crosses), and $d=3$ (small squares). The logarithmic scale for
the ordinate axis was used to visualize the structure of $\mathcal C(a)$.
    }
\end{center}
\end{figure}

To check whether $D(a)$ is uniformly anti-Holderian we concentrated on the case
$\gamma(a) = 1$, i.e. $c_1^* = 0$. Using a computer program we found
\emph{all} Markov slopes $a\in(2;8]$ that are algebraic numbers of degree $d
\le 4$ and correspond to $c_1^* = 0$. Then, for each such a slope, we
calculated the coefficient $\mathcal C(a)$. To this end we assumed that at each
investigated slope the graph of $\mathrm{osc}_\tau(D;a)$ as a function of
$\log\tau^{-1}$ behaves like the one depicted in \fref{Fig3}. To increase
accuracy we eliminated local periodic oscillations of the curves by using
\eref{def:lambda}. As expected, the points of each curve at abscissas forming
an arithmetic sequence with a common difference $\log\lambda$ turned out to
form practically a straight line with the slope quickly converging; we
identified this limiting slope with $\mathcal C(a)$. In each case the sum of
squares of the residuals from the best-fit line was less than $10^{-10}$. The
results obtained in this way are depicted in \fref{Fig4}a. As expected, the
density of the points increases with $a$. Moreover, apparently they lie  on  a
\emph{continuous} curve that crosses the $x$-axis only at $a=2$. These
findings are supported by the results depicted in the inset of this figure and
obtained for all Markov slopes $a\in(2;3]$ that correspond to $c_1^* = 0$ and
$d \le 6$. That $\mathcal C(a)$ tends to a continuous curve implies that the
graph of $D$ as a function of $a$ is uniformly anti-Holderian with $H=1$ and
$\gamma = 1$ at least on the set of all Markov slopes $a>2$ for which $c_1^* =
0$. Note that this set is dense on $(2,\infty)$.

To check whether $D(a)$ is uniformly Holderian we focused on the Markov
slopes with  $\gamma(a) = 2$, i.e., $c_1^* \neq 0$. We investigated
numerically all Markov slopes $a\in(2;8]$ that are algebraic numbers of degree
$d \le 3$ and correspond to $c_1^* \neq 0$ (286 data points). We found that at
each such a slope the exponent $\gamma$ is actually equal to 2. However, as
seen in \fref{Fig4}b, the coefficient $\mathcal C(a)$ turned out to be a
highly irregular, discontinuous function of $a$. Fortunately, this function is
apparently bounded from above. Taking into account our earlier result for
$c_1^* = 0$, we conclude that the graph of the diffusion coefficient $D$ as a
function of the slope $a$ is uniformly Holderian (with $H=1$ and $\gamma
= 2$) at least on the (dense) set of all Markov slopes.

The analysis of the H\"older and anti-H\"older condition at non-Markov slopes
is far more difficult.
Such values of $a$ were already investigated
numerically in \cite{Koza04}. The conclusion was that in this case most
probably $\gamma =1$, but this property cannot be definitely established
numerically due to very poor convergence. In several cases there is a clear
linear trend in graphs of  $\mathrm{osc}_\tau(D;a)$ as a function of
$\log\tau^{-1}$, which permits to estimate the value of $\mathcal C(a)$
(assuming $\gamma(a) =1$). For example, using the data of \cite{Koza04},
 we can estimate $\mathcal C(\pi) \approx
0.215$ and $\mathcal C(4.5) \approx 0.15$. As seen in \fref{Fig4}, these
values are very close to those obtained for the nearby Markov slopes. However,
for many other values of $a$ no reasonable numerical estimation of $\mathcal
C(a)$ is possible.

\subsection{Oscillations of $N_{k,l}(a)$}
\label{subsub:oscN}
Results of \sref{sub:bar.Xi} reveal that the logarithmic corrections to
fractal dimensions of graphs of transport coefficients must be attributed to
similar logarithmic corrections in functions $N_{k,l}(a)$. For $a\in\tilde\Xi$
these latter functions are continuous in $a$ and we expect them to satisfy
\eref{conjecture} with $N_{k,l}$ substituted for $D$, i.e., for $\tau \to 0$
\begin{equation}
 \label{conjecture:Nkl}
   \frac{\textrm{osc}_\tau(N_{k,l};a)}{\tau} \sim
  \left[\log(\tau^{-1})\right]^{\gamma_{k,l}(a)}
\end{equation}
 where, following \eref{eq:gen.osc.bound.1} and
\eref{eq:gen.osc.bound.2},
\begin{eqnarray}
  \label{eq:gamma.kl.1}
  0 \le \gamma_{k,l} \le k + l -1 &\quad&\textrm{for } c_1^*>0,\\
  \label{eq:gamma.kl.2}
  0 \le \gamma_{k,l}\le l &\qquad&\textrm{for } c_1^*=0.
\end{eqnarray}
This conjecture relates the fractal nature of transport coefficients to
fractality of functions $N_{k,l}(a)$ which---although of no immediate physical
meaning---have much simpler form and are more amenable to rigorous treatment.

To verify \eref{conjecture:Nkl} -- \eref{eq:gamma.kl.2} we have calculated
oscillations of $N_{3,0}$ and $N_{1,1}$ for the vanishing bias $b$ and two
values of the slope $a$: the largest root of   $a^5 - 4a^4 + 2a^3 + 3a^2 - 2a
+ 4$ (i.e., for $a\approx 3.04$) and the largest root of $a^5 - 4a^4 - 4a^2 + a
+ 4$ (i.e., for  $a\approx 4.20$). For these parameter values the system has a
 Markov partition \cite{Koza04}; moreover, the constant $c^*_1=0$ for
$a\approx 3.04$ and $c^*_1>0$ for $a\approx 4.20$.

\begin{figure}
  \includegraphics[width=\columnwidth, clip=true]{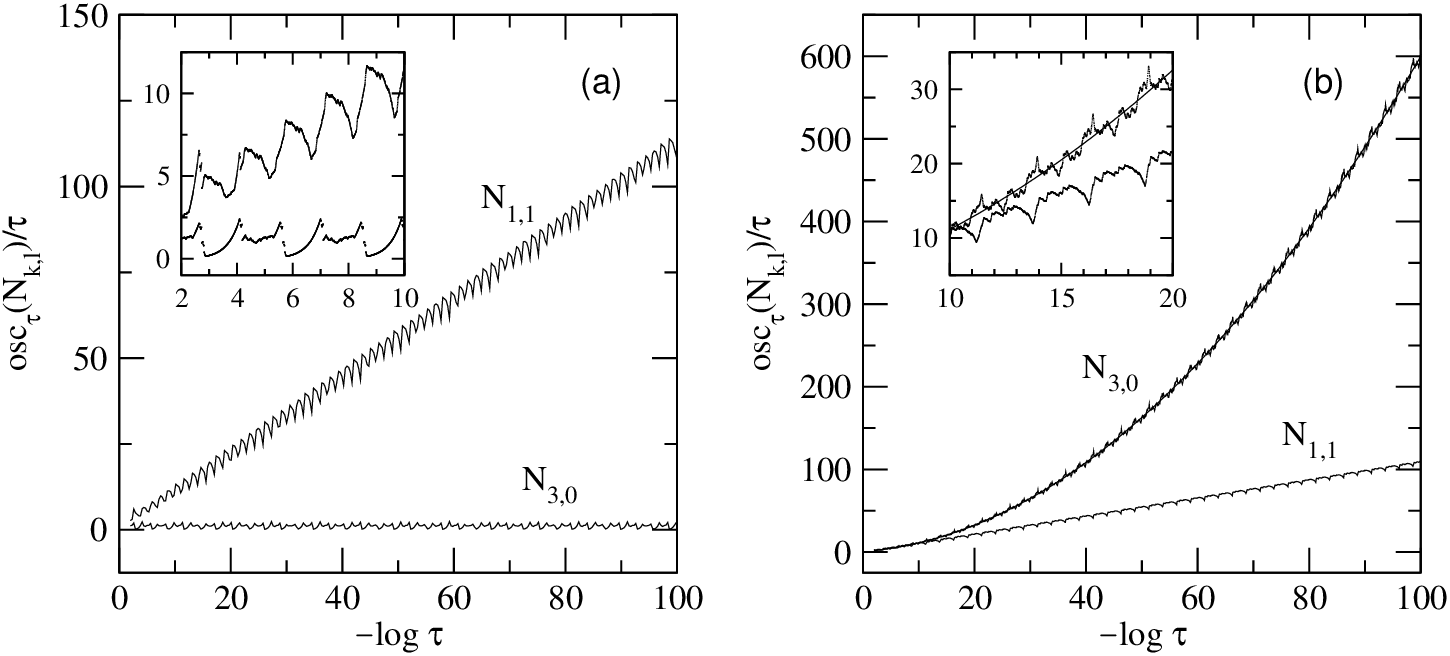}
  \caption{
    \label{Fig5}
    Oscillations of $N_{1,1}$ and $N_{3.0}$ rescaled in accordance with
    (\protect\ref{conjecture:Nkl}) calculated on intervals
    $(a-\tau,a+\tau)$ for $10^{-100} \le \tau \le 10^{-2}$, with $b=0$ and
    $a$ equal to the largest root of  (a) $a^5 - 4a^4 + 2a^3 + 3a^2 - 2a + 4$,
    (b)~$a^5 - 4a^4 - 4a^2 + a + 4$. Results for $N_{3,0}$ in the right panel were
    fitted by a quadratic (a solid line, hardly distinguishable from numerical
    data). The insets are blowups of the curves to show their periodicity and
    discontinuity in $a$.
 }
\end{figure}

Our results for $a\approx 3.04$ and $a\approx 4.20$ are depicted in figures
\ref{Fig5}a and \ref{Fig5}b, respectively. As we can see, our data clearly
indicate that in both cases $\gamma_{1,1} = 1$. However, $\gamma_{3,0} = 0$ in
\fref{Fig5}a (i.e., for $c^*_1=0$) and  $\gamma_{3,0} = 2$ in \fref{Fig5}b
(i.e., for $c^*_1>0$). All these results are in perfect agreement with
\eref{eq:gamma.kl.2} and actually correspond to the upper bounds implied by
these equations. Note that for $c^*_1=0$ equation \eref{eq:gamma.kl.2} actually
gives not only the upper bound, but the exact value of $\gamma_{k,0} =0$ for
all $k>0$.

The insets in \fref{Fig5} suggest that, just as in the case of graphs of
$\textrm{osc}_\tau(D)/\tau$ discussed in \sref{subsub:c}, on finer scales the
graphs of $\textrm{osc}_\tau(N_{k,l})/\tau$ as functions of $\log\tau^{-1}$
have periodic components. Their period lengths $\lambda$ can be calculated
using \eref{m-limit}, which gives $\lambda= a^6 \approx 794$ for $a\approx
3.04$ (\fref{Fig5}a) and $\lambda= a^4 \approx 311$ for $a\approx 4.20$
(\fref{Fig5}b). These theoretical values are in excellent agreement with
numerical data used to produce \fref{Fig5}.

\subsection{Self-similarity of the graphs}
\label{sub:self-similarity}

\begin{figure}
  \includegraphics[width=\columnwidth, clip=true]{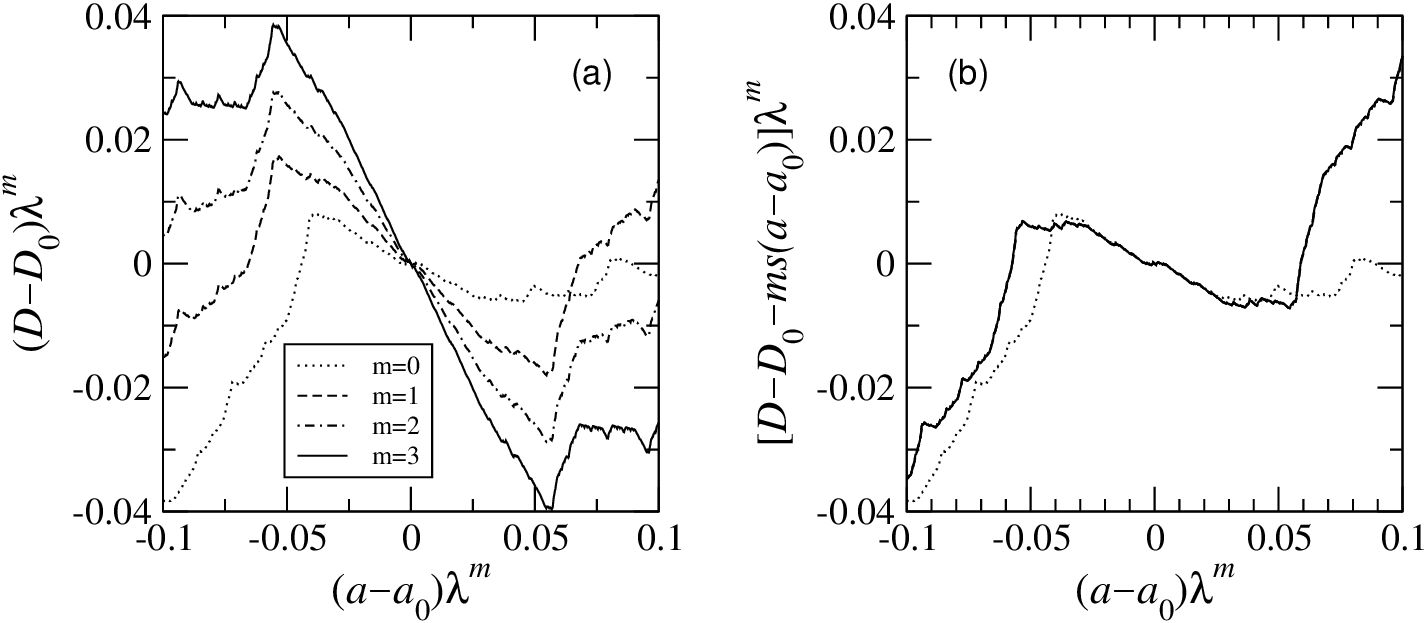}
  \caption{
    \label{Fig6}
    (a) The graph of the diffusion coefficient obtained for the
    slope $a_0\approx 2.455$ and the bias $b=0$ (dotted line) and its
    consecutive $\lambda$-fold magnifications about the point $(a_0, D_0)$,
    where $\lambda = a_0^5\approx 89$ and $D_0\equiv D(a_0)\approx 0.0967$.
    (b)~The same as in (a), except that now each $\lambda$-fold magnification
    has been followed by a linear transformation $(x,y) \mapsto (x,y-sx)$
    with $s\approx -0.1969$; all 4 curves have been plotted, but those
    obtained for $m=1,2,3$ differ from each other by less than the line width.
    }
\end{figure}

In the limit of $\tau\to0$ the periodicity of the curves presented in figures
\ref{Fig3} and \ref{Fig5} becomes practically perfect.  This phenomenon could
be a consequence of local self-similarity of functions $D(a)$ and $N_{k,l}(a)$
under enlargement of their graphs by a scale factor $\lambda$. To verify this
hypothesis in \fref{Fig6}a we have plotted the graph of the diffusion
coefficient $D$ as a function of the slope $a$, with the origin moved to the
point $a_0\approx 2.455$ and $D_0 \equiv D(a_0)\approx 0.0967$. Then we
plotted three consecutive blowups of this graph obtained by magnifying it
about the new origin by scale factors $\lambda$, $\lambda^2$, $\lambda^3$,
where $\lambda = a_0^5\approx 89$. As we can see in \fref{Fig6}a, this simple
transformation of graphs does not lead to perfectly self-similar structures;
nevertheless, the curves thus obtained do exhibit some kind of similarity.

The effect of actual self-similarity can be achieved if each $\lambda$-fold
enlargement will be followed by a correction transformation related
to the nonvanishing logarithmic correction exponent $\gamma$. The 2-step
transformation $T$ can be expressed as
\begin{equation}
  \label{eq:T}
   T(x, y) = (\lambda x, \lambda y - \kappa_1(\lambda  x)),
\end{equation}
where $\lambda$ is the scale factor and $\kappa_1$ is a correction term. Our
numerical results suggest that for $\gamma=1$ this complementary step is a
simple linear transformation, $\kappa_1(x) = sx$, with $s$  being a constant
depending on the slope $a$, the bias $b$ and the function the self-similarity
transformation $T$ is applied to. In the case of the data shown in \fref{Fig6}
the constant $s \approx -0.1969$. The net effect of this two-step scaling
procedure applied iteratively $m=0,1,2,3$ times is depicted in \fref{Fig6}b.
As we can see, the convergence is very fast and the curves obtained after $m=$
1, 2 and 3 elementary 2-step transformations are already practically
indistinguishable from each other. Closer analysis of the numerical data for
$m=0,\ldots,10$ revealed that with each 2-step transformation we get closer to
the self-similar limit curve by about two significant figures. Similar
behaviour was found for several other points $a_0$, both for graphs of the
diffusion coefficient $D(a)$ and the auxiliary functions $N_{1,1}(a)$,
$N_{3,0}(a)$, provided that the corresponding logarithmic correction exponent
$\gamma$ was 1. Note that transformation of type (\ref{eq:T}) holds also for
Takagi functions.

\begin{figure}
  \includegraphics[width=\columnwidth, clip=true]{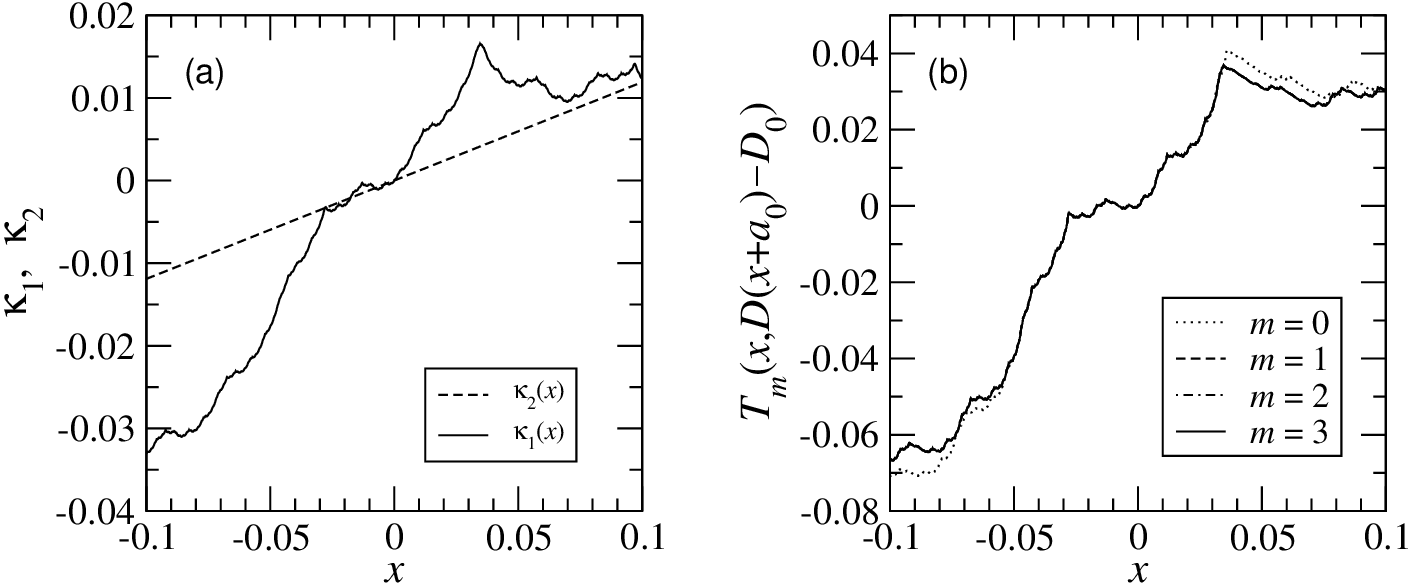}
  \caption{
    \label{Fig7}
    (a) Correction functions $\kappa_1$ and $\kappa_2$ used in self-similarity
    transformation \protect\eref{eq:T.m.2} for the graph of the diffusion
    coefficient $D(a)$ about the point ($a_0, D_0)$ with $a_0 \approx 4.20$,
    $D_0=D(a_0)$, and the bias $b=0$. (b)~The self-similarity effect of
    applying $T_m$, as described by \protect\eref{eq:T.m.2}, to the graph of
    $D(a)$, for the same parameter values as those used to generate panel (a).
    Note that the dotted line (corresponding to $m=0$) actually represents the
    graph of the original function $D(a)$ translated by a vector $(-a_0,
    -D_0)$ and $x$ can be interpreted as $\lambda^m(a-a_0)$. The graphs
    obtained for $m\ge 1$ are practically indistinguishable from each other. }
\end{figure}

For $\gamma =1 $ the net effect of applying consecutively $m$ transformation
$T$ is
\begin{equation}
  \label{eq:T.m}
   T_m(x, y) = (\lambda^m x, \lambda^m y - m\kappa_1(\lambda^m  x)),
\end{equation}
where linearity in $m$ of the correction term corresponds to $\gamma=1$. For
$\gamma=2$ the situation gets more complicated. We now expect that the
correction term after $m$ self-similarity transformations is quadratic in $m$,
\begin{equation}
  \label{eq:T.m.2}
   T_m(x, y) =
   (\lambda^m x, \lambda^m y -
       m^2\kappa_2(\lambda^m  x) -
       m\kappa_1(\lambda^m x)),
\end{equation}
with some real functions $\kappa_1$ and $\kappa_2$. A side-effect of such a
form of the correction term is that, in contrast to the case $\gamma=1$, now
the second step in each elementary 2-step transformation depends on $m$ and
hence slightly differs from each other. We have confirmed \eref{eq:T.m.2}
numerically, finding the convergence to be very fast. In particular,
\fref{Fig7} presents results obtained for the slope $a_0\approx 4.20$ and bias
$b_0=0$ (these parameters were already used in \fref{Fig5}b). On the left
panel we show correction functions $\kappa_1$ and $\kappa_2$. As we can see,
$\kappa_2$ is linear, while $\kappa_1$ is an irregular (perhaps fractal)
function of its argument. Although the graph of $\kappa_1(x)$ in \fref{Fig7}a
is quite similar to that of $D(a)$ (dotted line in \fref{Fig7}b), these two
functions are different. On the right panel we present effects of applying
$T_m$, $m=0,1,2,3$, to the graph of $D(a)$ in the vicinity of the point $(a_0,
D(a_0))$. The convergence to the self-similar limit is very fast. Actually for
$m>1$ the graphs of $T_m$ in \fref{Fig7}b differ from each other by less than
the line width.

\section{Summary and conclusions}
\label{sec:Conclusions}

We have analysed both analytically and numerically the local
(Minkowski-Bouligand) fractal dimension of graphs of transport coefficients as
functions of the slope $a$ in a simple dynamical system defined by a
piece-wise linear map $M_{a,b}$. We showed that for all values of the control
parameters $a$ and $b$ such that the system defined by $M_{a,b}$ has a Markov
partition, the dimension of such graphs is 1, probably with a logarithmic
correction. This correction renders the curves nowhere rectifiable (i.e., any
of their arcs is ``of infinite length''), which is typical of fractals.
 Our results imply also that the Hausdorff dimension of these
curves is 1, too.

We have also found the upper bound for the exponent $\gamma$ controlling the
logarithmic  correction. This bound turned out to be either 1 or 2, depending
on whether a constant $c^*_1$ vanishes or not, or, equivalently, whether the
trajectories of $x_0 = \pm\half$ remain bounded or not. This finding extends
our earlier conjecture \cite{Koza04} about factors determining the value of
$\gamma$.

Using numerical calculations we have shown that at Markov slopes $a$ the graph
of the diffusion coefficient is actually self-similar under the action of a
special, two-step scaling  transformation. The first step is a simple uniform
enlargement by a scale factor $\lambda$. This transformation must be
followed by the second step, a ``correction''  transformation of the graph,
which is related to the non-zero value of the exponent $\gamma$  controlling
the logarithmic correction to the local fractal dimension. We found the
precise formula for the value of the scale factor $\lambda$ as well as a
general form of the correction transformations.  In particular, we found that
the scale factor $\lambda$ depends on the slope $a$ and the periodicity of the
trajectories of points $x_0=\pm\half$. The correction transformations turned
out linear (affine) for $\gamma = 1$ and different, more complicated for
$\gamma=2$. At non-Markov slopes the graphs are probably not
self-similar.

The graphs of transport coefficient have thus surprisingly reach spectrum of
local fractal properties. Our numerical results suggest that the curves are
self-similar on a dense set of points (at which the system has a Markov
partition), but at each such a point the scale factor and the correction
transformation are different. In this sense the curves are multifractals.

Our  analytical approach has, however, some limitations. The most serious one
is related to the fact that it is applicable only to  systems in which the
trajectories of points $x_0=\pm\half$ are periodic (or, equivalently, to
systems with a Markov partition). We believe that the typical value of
$\gamma$ at non-Markovian slopes $a$ is 1, but in general this value can take
on any value between 1 and 2. This conjecture, however, requires further
studies.
  Proving that the investigated curves are \emph{uniformly} Holderian with $H=1$ and $\gamma=2$ and
  anti-Holderian with $H=1$ and $\gamma=1$ is another open problem. So far we have been able to justify
  these properties only numerically.

Numerical results suggest that our analytical approach gives not only the
upper bound, but  the exact value of $\gamma$. To prove this hypothesis
rigorously would require finding a ``tight'' lower bound for $\gamma$.
However, as is well known \cite{Tricot}, finding the lower bound of a fractal
dimension is usually much more difficult that finding the upper one. This is
also true in our case. For example, when we were looking for the upper bound
of oscillations of the quotient of two functions, we did not have to take into
account the fact that actually their values were correlated. Applying the same
general assumption for the lower bound of $\gamma$ yields only a trivial result
$\gamma \ge 0$. Therefore finding a better lower bound for the logarithmic
correction is another problem for future studies.

Another  question is how to reconcile our results with those obtained by
Klages and Klau\ss\ \cite{KlagesKlauss03}, who analysed the box-counting
 dimension of $D(a)$ on intervals of \emph{finite} length and found it
to be greater than 1. One possible way of answering this problem could be to
analyse how the local self-similarity of the curves depends on the box size.

Although most of our analysis was carried out explicitly for the graph of the
diffusion coefficient $D$ as a function of the slope $a$ for a vanishing bias
$b$, we showed how it can be generalized for other transport coefficients and
for arbitrary bias $b$. Similarly we expect that the main conclusions derived
here for the simple piece-wise linear map $M_{a,b}$ remain valid also in the
case of more realistic (and complex) dynamical systems. After all, in our
model the main reason for irregular, fractal dependence of the transport
coefficients on the slope $a$ is an extreme sensitivity of the (periodic)
Markov orbits on the control parameters; similar sensitivity is observed in
more complex models, e.g.\ the multiBaker or the Lorentz gas
\cite{Har02,Lloyd95,GaspardKlages98,KlagesHab,Vollmer03}. Transport
coefficients in these models were already called ''fractal'', but only in the
sense of ''unexpectedly irregular''. We hope that our present approach will
help justifying a conjecture that the transport coefficients in those more
realistic models form fractals in a strictly mathematical sense of this term.

\ack I thank R.\ Klages for introducing me into the subject, many invaluable remarks on the manuscript and inspiration to write section \ref{subsub:plot-of-c}. Support from the Polish KBN Grant Nr 2 P03B 030 23 is also gratefully
acknowledged.

\end{document}